\documentclass[journal]{IEEEtran}

\usepackage{cite}

\ifCLASSINFOpdf
  \usepackage[pdftex]{graphicx}
  \graphicspath{{../pdf/}{../jpeg/}}
   \DeclareGraphicsExtensions{.pdf,.jpeg,.png}
\else
  \usepackage[dvips]{graphicx}
  \graphicspath{{../eps/}}
   \DeclareGraphicsExtensions{.eps}
\fi
\usepackage[cmex10]{amsmath}
\usepackage{array}

\ifCLASSOPTIONcompsoc
  \usepackage[caption=false,font=normalsize,labelfont=sf,textfont=sf]{subfig}
\else
 \usepackage[caption=false,font=footnotesize]{subfig}
\fi

\usepackage{amsmath,graphicx}
\usepackage{amssymb}
\usepackage{amscd}

\def\expk{{e^{-i \frac{2\pi}{T}k t}}}

\begin{document}
%
\title{Fourier Domain Beamforming:\\
The Path to Compressed Ultrasound Imaging}
%
%
%

\author{Tanya Chernyakova,~\IEEEmembership{Student Member,~IEEE,}
        Yonina C. Eldar,~\IEEEmembership{Fellow,~IEEE}}

\maketitle

\begin{abstract}
Sonography techniques use multiple transducer elements for tissue visualization. Signals detected at each element are sampled prior to digital beamforming. The sampling rates required to perform high resolution digital beamforming are significantly higher than the Nyquist rate of the signal and result in considerable amount of data, that needs to be stored and processed. A recently developed technique, compressed beamforming, based on the finite rate of innovation model, compressed sensing (CS) and Xampling ideas, allows to reduce the number of samples needed to reconstruct an image comprised of strong reflectors. A drawback of this method is its inability to treat speckle, which is of significant importance in medical imaging.
Here we build on previous work and extend it to a general concept of beamforming in frequency. This allows to exploit the low bandwidth of the ultrasound signal and bypass the oversampling dictated by digital implementation of beamforming in time.
Using beamforming in frequency, the same image quality is obtained from far fewer samples.
We next present a CS-technique that allows for further rate reduction, using only a portion of the beamformed signal's bandwidth. We demonstrate our methods on in vivo cardiac data and show that reductions up to 1/28 over standard beamforming rates are possible.
Finally, we present an implementation on an ultrasound machine using sub-Nyquist sampling and processing. Our results prove that the concept of sub-Nyquist processing is feasible for medical ultrasound, leading to the potential of considerable reduction in future ultrasound machines size, power consumption and cost.\end{abstract}

\begin{IEEEkeywords}
Array processing, beamforming, compressed sensing (CS), ultrasound, sub-Nyquist sampling.
\end{IEEEkeywords}

%
\IEEEpeerreviewmaketitle

\section{Introduction}

%
%
%
%

%

\label{sec:intro}
Diagnostic ultrasound has been used for decades to visualize body structures. Imaging is performed by transmitting a pulse along a narrow beam from an array of transducer elements. During its propagation echoes are scattered by acoustic impedance perturbations in the tissue, and detected by the array elements. The data, collected by the transducers, is sampled and digitally integrated in a way referred to as beamforming, which results in signal-to-noise ratio (SNR) enhancement and improvement of angular localization. Such a beamformed signal, referred to as beam, forms a line in the image.

According to the classic Shannon-Nyquist theorem \cite{shannon1949communication}, the minimal sampling rate at each transducer element should be at least twice the bandwidth of the detected signal in order to avoid aliasing. In practice, rates up to $4$-$10$ times the central frequency of the transmitted pulse are required in order to eliminate artifacts caused by digital implementation of beamforming in time\cite{steinberg1992digital}.
Taking into account the number of transducer elements and the number of lines in an image, the amount of sampled data that needs to be digitally processed is enormous, motivating methods to reduce sampling and processing rates. 

A possible approach to sampling rate reduction is introduced in \cite{tur2011innovation}. Tur et al. consider the ultrasound signal detected by each receiver within the framework of finite rate of innovation (FRI) \cite{vetterli2002sampling}. The detected signal is modeled as $L$ replicas of a known pulse-shape, caused by scattering of the transmitted pulse from various reflectors, located along the transmitted beam. Such an FRI signal is fully described by $2L$ parameters, corresponding to the replica's unknown delays and amplitudes. Based on \cite{vetterli2002sampling}, the relationship between the signal's Fourier series coefficients and the unknown parameters is formulated in the form of a spectral analysis problem. The latter may be solved using array processing methods or compressed sensing (CS) techniques, given a subset of at least $2L$ Fourier series coefficients \cite{wagner2012compressed}, \cite{eldar2012compressed}. The required Fourier coefficients can be computed from appropriate low-rate samples of the signal following ideas of \cite{tur2011innovation, michaeli2011xampling, mishali2011xampling, mishali2011xampling2, gedalyahu2011multichannel}. Recent work has developed a hardware prototype implementing the suggested sub-Nyquist system \cite{Baransky2012radar}.

The above framework allows to sample the detected signals at a low-rate, assuming sufficiently high SNR.  However, the final goal in low-rate ultrasound imaging is to recover a two-dimensional image, obtained by integrating the noisy data sampled at multiple transducer elements. In standard imaging the integration is achieved by the process of beamforming, which is performed digitally and, theoretically, requires high sampling rates.
Hence, in order to benefit from the rate reduction achieved in \cite{tur2011innovation}, one needs to be able to incorporate beamforming into the low-rate sampling process.

\subsection{Related Work: Compressed Beamforming}
\label{ssec:Compressed BF}
A solution to low-rate beamforming is proposed in \cite{wagner2012compressed}, where Wagner et al. introduce the concept of compressed beamforming. They show that their approach, applied to an array of transducer elements, allows to reconstruct a two-dimensional ultrasound image depicting macroscopic perturbations in the tissue.
To develop their method, the authors first prove that the beam obeys an FRI model, implying that it can be reconstructed from a small subset of its Fourier coefficients. However, this required subset cannot be obtained by the schemes proposed in \cite{tur2011innovation} and \cite{gedalyahu2011multichannel}, since the beam does not exist in the analog domain. It is constructed digitally after sampling the detected signals. This fundamental obstacle is resolved by transforming the beamforming operator into the compressed domain. Specifically, Wagner et al. show that the Fourier coefficients of the beam can be approximated by a linear combination of Fourier coefficients of the detected signals. The latter are obtained from the low-rate samples of the detected signals, using the Xampling method, proposed in  \cite{tur2011innovation}, \cite{gedalyahu2011multichannel} and \cite{Baransky2012radar}.

Another innovation of \cite{wagner2012compressed} regards the approach to beam reconstruction from a subset of its frequency samples. Rather than use standard spectral analysis techniques, Wagner et al. view the reconstruction as a CS problem. They demonstrate that CS methodology is comparable to spectral analysis methods and even outperforms the latter when the noise is large.
Combining compressed beamforming with CS techniques for signal recovery, they reconstruct two-dimensional ultrasound images, comprised of strong reflectors in the tissue. Significant rate reduction is achieved, while assuming that the number of replicas in the FRI model of the beam is small. Such an assumption is justified by the fact that only strong perturbations in the tissue are taken into account. Therefore, the proposed framework allows for robust detection of strong reflectors, but is unable to treat speckle, weak scattered echoes originating from microscopic perturbations in the tissue, which are of significant importance in medical imaging.

%

\subsection{Contributions}
\label{ssec:Contributions}
In this paper we build on the results in \cite{wagner2012compressed} and show that compressed beamforming can be extended to a much more general concept of beamforming in frequency. This approach to beamforming is applicable to any signal, without the need to assume a structured model. When structure exists, beamforming in frequency may be combined with CS to yield further rate reduction.

The core of compressed beamforming is the relationship between the beam and the detected signals in the frequency domain, while the notion of ``compressed'' stems from the fact that the Fourier coefficients of the detected signals can be obtained from their low-rate samples. Here we show that this frequency domain relationship is general and holds irrespective of the FRI model. This leads to an approach of beamforming in frequency which is completely equivalent to beamforming in time.
Beamforming in frequency is equivalent to a weighted averaging of the Fourier coefficients of the detected signals and can be performed efficiently by exploiting two facts. First, the frequency domain beamforming operator is defined by the geometry of the transducer array and does not depend on the detected signals. Hence, the required weights can be computed off-line and used as a look-up-table during the imaging cycle. In addition, we show numerically, that these weights are characterized by a rapid decay, implying that the Fourier coefficients of the beam can be computed using a small number of Fourier coefficients of the detected signals.

Next, we show that beamforming in frequency domain allows to bypass the oversampling dictated by digital implementation of beamforming in time. Since the beam is obtained directly in frequency, we need to compute its Fourier coefficients only within its effective bandwidth. We demonstrate that this can be achieved using generalized samples of the detected signals, obtained at their Nyquist rate.
To avoid confusion, by Nyquist rate we mean the signals effective bandpass bandwidth, which is typically much lower than its highest frequency since the detected pulse is normally modulated onto a carrier and only occupies a portion of the entire bandwidth.
Using in vivo cardiac data, we illustrate that beamforming in frequency allows to preserve image integrity with $4$-$10$ fold reduction in the number of samples used for its reconstruction.

Further reduction in sampling rate is obtained, similarly to \cite{wagner2012compressed}, when only a portion of the beam's bandwidth is used. In this case beamforming in frequency is equivalent to compressed beamforming. Detected signals are sampled at sub-Nyquist rates, leading to up to $28$ fold reduction in sampling rate. Our contribution in this scenario regards the reconstruction method used to recover the beam from its partial frequency data.
To recover the unknown parameters, corresponding to the FRI model of the beam, Wagner et al. assume that the parameter vector is sparse. The parameters are then obtained as a solution of an $\textit{l}_0$ optimization problem. Sparsity holds when only strong reflectors are taken into account, while the speckle is treated as noise.
To capture the speckle, we assume that the parameter vector is compressible and recast the recovery as an $\textit{l}_1$ optimization problem. We show that these small changes in the model and the CS reconstruction technique allow to capture and recover the speckle, leading to significant improvement in image quality.

Finally, we introduce an implementation of beamforming in frequency and sub-Nyquist processing on a stand alone ultrasound machine and show that our proposed processing is feasible in practice using real hardware. Low-rate processing is performed on the data obtained in real-time by scanning a heart with a $64$-element probe. Our approach allows for significant rate reduction with respect to the lowest processing rates that are achievable today, which can potentially impact system size, power consumption and cost.
\\

The rest of the paper is organized as follows: in Section \ref{sec:conv processing}, we review beamforming in time and discuss the sampling rates required for its digital implementation. Following the steps in \cite{wagner2012compressed}, we describe the principles of frequency domain beamforming in Section \ref{sec:beam in freq}, and show that it is equivalent to standard time domain processing. In Section \ref{sec:rate reduction} we show that beamforming in frequency allows for rate reduction even without exploiting the FRI model and can be performed at the Nyquist rate of the signal. CS recovery from partial frequency data, implying sampling and processing at sub-Nyquist rates, is discussed in Section \ref{sec:CS}.
Comparison between the performance of the proposed method with the results obtained in \cite{wagner2012compressed} together with an implementation of beamforming in frequency and sub-Nyquist processing on a stand alone ultrasound machine are presented in Section \ref{sec:simulations and results}.

\section{Conventional Processing in Ultrasound Imaging}
\label{sec:conv processing}
Most modern imaging systems use multiple transducer elements to transmit and receive acoustic pulses. This allows to perform beamforming during both transmission and reception. Beamforming is a common signal-processing technique \cite{van2004detection} that enables spatial selectivity of signal transmission or reception and is applied in various fields, including wireless communication, speech processing, radar and sonar.
In ultrasound imaging beamforming is used for steering the beam in a desired direction and focusing it in the region of interest in order to detect tissue structures.

During transmission beamforming is achieved by delaying the transmission time of each transducer element, which allows to transmit energy along a narrow beam. Beamforming upon reception is much more challenging. Here dynamically changing delays are applied on the signals detected at each one of the transducer elements prior to averaging. Time-varying delays allow dynamic shift of the reception beam's focal point, optimizing angular resolution. Averaging of the delayed signals in turn enhances the SNR of the resulting beamformed signal, which is used to form a line in an image.
From here on, the term beamforming will refer to beamforming on reception, which is the focus of this work.

\subsection{Beamforming in Time}
\label{ssec:BF in time}
We begin with a detailed description of the beamforming process which takes place in a typical B-mode imaging cycle. Our presentation is based mainly on \cite{jensen1999linear} and \cite{wagner2012compressed}. We will then show, in Section \ref{sec:beam in freq}, how the same process can be performed in frequency, paving the way to substantial rate reduction.
\begin{figure}[htb]
\begin{minipage}[b]{1.0\linewidth}
  \centering
\vspace{-0.5cm}
  \centerline{\includegraphics[width=7cm]{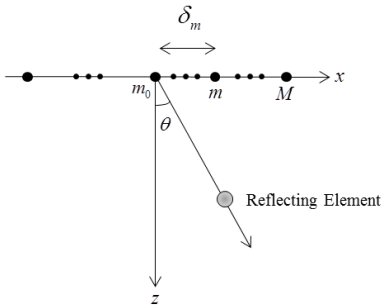}}
\end{minipage}
\caption{$M$ receivers aligned along the $x$ axis. An acoustic pulse is transmitted in a direction $\theta$. The echoes scattered from perturbation in the radiated tissue are received by the array elements.}
\label{fig:array}
\end{figure}

In the transmit path, a pulse is generated and transmitted by the array of transducer elements. The pulse transmitted by each element is timed and scaled, so that the superposition of all transmitted pulses creates a directional beam propagating at a certain angle. By subsequently transmitting at different angles, a whole sector is radiated. The real time computational complexity in the transmit path is negligible since transmit parameters per angle are calculated off-line and saved in tables.

Consider an array comprised of $M$ transceiver elements aligned along the $x$ axis, as illustrated in Fig. \ref{fig:array}. The reference element $m_0$ is set at the origin and the distance to the $m$th element is denoted by $\delta_m$. The image cycle begins at $t=0$, when the array transmits an energy pulse in the direction $\theta$. The pulse propagates trough the tissue at speed $c$, and at time $t\geq0$ its coordinates are $(x,z)=(ct\sin{\theta},ct\cos{\theta})$. A potential point reflector located at this position scatters the energy, such that the echo is detected by all array elements at a time depending on their locations. Denote by $\varphi_m(t;\theta)$ the signal detected by the $m$th element and by $\hat{\tau}_m(t;\theta)$ the time of detection. It is readily seen that:
\begin{equation}
\label{eq:tau^m}
\hat{\tau}_m(t;\theta)=t+\frac{d_m(t;\theta)}{c},
\end{equation}
where $d_m(t;\theta)=\sqrt{(ct\cos{\theta})^2+(\delta_m-ct\sin{\theta)}^2}$ is the distance traveled by the reflection.
Beamforming involves averaging the signals detected by multiple receivers while compensating the differences in detection time. In that way we obtain a signal containing the intensity of the energy reflected from each point along the central transmission axis $\theta$.

Using \eqref{eq:tau^m}, the detection time at $m_0$ is $\hat{\tau}_{m_0}(t;\theta)=2t$ since $\delta_{m_0}=0$. Applying an appropriate delay to $\varphi_m(t;\theta)$, such that the resulting signal $\hat{\varphi}_m(t;\theta)$ satisfies $\hat{\varphi}_m(2t;\theta)=\varphi_m(\hat{\tau}_m(t;\theta))$, we can align the reflection detected by the $m$-th receiver with the one detected at $m_0$. Denoting $\tau_m(t;\theta)=\hat{\tau}_m(t/2;\theta)$  and using \eqref{eq:tau^m}, the following aligned signal is obtained:
\begin{align}
\label{eq:phim}
\hat{\varphi}_m(t;\theta)&=\varphi_m(\tau_m(t;\theta);\theta),\\ \nonumber
\tau_m(t;\theta)&=\frac{1}{2}\left(t+\sqrt{t^2-4(\delta_m/c)t\sin{\theta}+4(\delta_m/c)^2}\right).
\end{align}
The beamformed signal may now be derived by averaging the aligned signals:
\begin{equation}
\label{eq:phi beamformed}
\Phi(t;\theta)=\frac{1}{M}\sum_{m=1}^M{\hat{\varphi}_m(t;\theta)}.
\end{equation}
Such a beam is optimally focused at each depth and hence exhibits improved angular localization and enhanced SNR.

Although defined over continuous time, ultrasound imaging systems perform the beamforming process in \eqref{eq:phim}-\eqref{eq:phi beamformed} in the digital domain: analog signals $\varphi_m(t;\theta)$ are amplified and sampled by an Analog to Digital Converter (ADC), preceded by an anti-aliasing filter.
We next discuss sampling and processing rates required to perform \eqref{eq:phi beamformed}.

\subsection{Rate Requirements}
\label{ssec:rate requirements}
Digital implementation of beamforming requires sampling the signals detected at the transducer elements and transmitting the samples to the processing unit.
The Nyquist rate, required to avoid aliasing, is insufficient for digital implementation of beamforming due to the high delay resolution needed. Indeed, in order to apply the delay defined in \eqref{eq:phim} digitally, detected signals need to be sampled on a sufficiently dense grid. Typically, the sampling interval is on the order of nanoseconds. Therefore, required sampling rates are significantly higher than the Nyquist rate of the signal and can be as high as hundreds of MHz \cite{oDonnell1990real}.

Due to the impracticality of this requirement, ultrasound data is sampled at lower rates, typically, on the order of tens of MHz. Fine delay resolution is obtained by subsequent digital interpolation. Interpolation beamforming allows to reduce the sampling rate at the cost of additional computational load required to implement the digital interpolation which effectively increases the rate in the digital domain. The processing, or more precisely, beamforming rate, remains unchanged as it is performed at the high digital rate.

Another common way to improve delay accuracy while reducing both sampling and beamforming rate is phase-rotation-based beamforming (PRBF) \cite{steinberg1992digital}. In this approach coarse delays, defined by the sampling rate, are followed by a vernier control, implemented by a digital phase shift, adjusted for the central frequency. The phase shifter approximation to a time delay is exact only at the central frequency, leading to loss in array gain and rise in the sidelobe level. The analysis in \cite{steinberg1992digital} shows that the degradation of beam quality can be avoided, provided that the sampling rate is $4$-$10$ times the transducer central frequency. This rule of thumb stems from the assumption that typically the transducer central frequency is approximately twice the radio frequency (RF) bandwidth. RF bandwidth is defined as the distance from the central to the highest frequency and, hence, is half the bandpass bandwidth. This leads to the conclusion that the sampling rate should be about $4$-$10$ times the bandpass bandwidth, since, according to the analysis in \cite{steinberg1992digital}, loss in array gain and rise in the sidelobe level are dictated by the ratio between the bandwidth of the signal to the sampling rate.
In the sequel, following \cite{steinberg1992digital}, we denote the rate required to avoid artifacts in digital implementation of beamforming, as the beamforming rate $f_s$.\\
$~~~$As imaging systems evolve, the amount of elements participating in the imaging cycle continues to grow significantly. Consequently large amounts of data need to be transmitted from the system front-end and digitally processed in real time. Increasing transmission and processing pose an engineering challenge on digital signal processing (DSP) hardware and motivate reducing the amounts of data as close as possible to the system front-end.\\
$~~~$To conclude this section we evaluate the sampling rates and the number of samples needed to be taken at each transducer element according to each one of the methods, described above. Our evaluation is based on the imaging setup typically used in cardiac imaging. We assume a breadboard ultrasonic scanner of 64 acquisition channels. The radiated depth $r=16$ cm and speed of sound $c=1540$ m/sec yield a signal of duration $T=2r/c\simeq210$ $\mu$sec. The acquired signal is characterized by a narrow bandpass bandwidth of $2$ MHz, centered at the carrier frequency $f_0\approx3.4$ MHz. In order to perform plain delay-and-sum beamforming with $5$ nsec delay resolution, detected signals should be sampled at the rate of $200$ MHz. Implementation of interpolation beamforming, used in many imaging systems, allows to reduce the sampling rate to $50$ MHz, while the required beamforming rate is obtained through interpolation in the digital domain. Hence, each channel yields $42000$ real valued samples, participating in beamforming.
Rates required by PRBF in this setup, vary from $8$ to $20$ MHz, leading to $1680$-$4200$ real valued samples, obtained at each transducer element.\\
$~~~$Evidently, processing in the time domain imposes high sampling rate and considerable burden on the beamforming block. We next show that the number of samples can be reduced significantly by exploiting ideas of sub-Nyquist sampling, beamforming in frequency and CS-based signal reconstruction.


%

\section{Beamforming in frequency}
\label{sec:beam in freq}
We now show that beamforming can be performed equivalently in the frequency domain, paving the way to substantial reduction in the number of samples needed to obtain the same image quality.
We extend the notion of compressed beamforming, introduced in \cite{wagner2012compressed}, to beamforming in frequency and show that a linear combination of the discrete Fourier transform (DFT) coefficients of the individual signals, sampled at the beamforming rate $f_s$,  yields the DFT coefficients of the beamformed signal, sampled at the same rate. This relationship is true irrespective of the signal structure.

\subsection{Implementation and Properties}
\label{ssec:imlementation of BF in freq}
We follow the steps in \cite{wagner2012compressed} and start from the computation of the Fourier series coefficients of the beam $\Phi(t;\theta)$.
As shown in \cite{wagner2012compressed}, the support of the beam $\Phi(t;\theta)$ is limited to $[0,T_B(\theta))$, where $T_B(\theta)<T$ and $T$ is defined by the transmitted pulse penetration depth. The value of $T_B(\theta)$ is given by \cite{wagner2012compressed}
\begin{equation} 
    T_B(\theta)=\min_{1\leq m \leq M}{\tau_m^{-1}(T;\theta)},
\end{equation}
where $\tau_m(t;\theta)$ is defined in \eqref{eq:phim}.
Denote the Fourier series coefficients of $\Phi(t;\theta)$ with respect to the interval $[0,T)$ by
\begin{equation}
\label{fourier coeff of beamformed 1}
c_k^s=\frac{1}{T}\int_0^T I_{[0,T_B(\theta))}(t) \Phi(t;\theta)\expk dt,
\end{equation}
where $I_{[a,b)}$ is the indicator function equal to $1$ when $a\leq t<b$ and $0$ otherwise.
Plugging \eqref{eq:phi beamformed} into \eqref{fourier coeff of beamformed 1}, and after some algebraic manipulation, it is shown in \cite{wagner2012compressed} that
\begin{equation}
\label{fourier coeff of beamformed 2}
c_k^s=\frac{1}{M}\sum_{m=1}^M c_{k,m}^s,
\end{equation}
where $c_{k,m}^s$ are defined as follows:
\begin{equation}
\label{c k m}
c_{k,m}^s=\frac{1}{T}\int_0^T g_{k,m}(t;\theta) \varphi_m(t;\theta)dt,
\end{equation}
with
\begin{align}\label{g j m q j m}
g_{k,m}(t;\theta)=&q_{k,m}(t;\theta)\expk, \nonumber \\
q_{k,m}(t;\theta)=& I_{[|\gamma_m|,\tau_m(T;\theta))}(t) \left(1+\frac{\gamma_m^2\cos{\theta}^2}{(t-\gamma_m\sin{\theta})^2}\right)\times  \\*
&\exp{\left\{i\frac{2\pi}{T}k\frac{\gamma_m-t\sin{\theta}}{t-\gamma_m\sin{\theta}}\gamma_m\right\}}, \nonumber
\end{align}
and $\gamma_m=\delta_m/c$.

The next step is to replace $\varphi_m(t)$ by its Fourier series coefficients. Denoting the $n$th Fourier coefficient by $\varphi_m^s[n]$ and using \eqref{g j m q j m} we can rewrite \eqref{c k m} as
\begin{equation}\label{c k m fourier}
 c_{k,m}^s=\sum_n \varphi_m^s[k-n]Q_{k,m;\theta}[n],
\end{equation}
where $Q_{k,m;\theta}[n]$ are the Fourier coefficients of the distortion function $q_{k,m}(t;\theta)$ with respect to $[0,T)$. According to Proposition 1 in \cite{wagner2012compressed}, $c_{k,m}^s$ can be approximated sufficiently well when we replace the infinite summation in \eqref{c k m fourier} by a finite sum:
\begin{equation}\label{c k m approx}
c_{k,m}^s\simeq\sum_{n\in\nu(k)}\varphi_m^s[k-n]Q_{k,m;\theta}[n].
\end{equation}
The set $\nu(k)$ depends on the decay properties of $\{Q_{k,m;\theta}[n]\}$.

\begin{figure}[htb]
\begin{minipage}[b]{1.0\linewidth}
  \centering
  \centerline{\includegraphics[width=6.5cm]{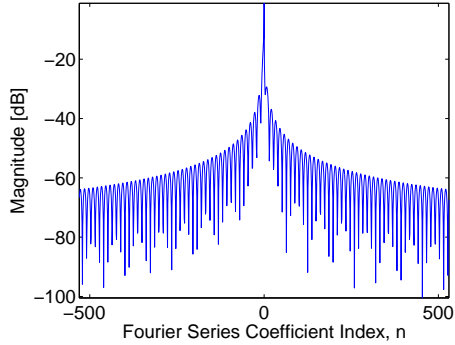}}
\end{minipage}
\caption{Fourier coefficients $\{Q_{k,m;\theta}[n]\}$ of $q_{k,m}(t;\theta)$ are characterized by a rapid decay, where most of the energy is concentrated around the DC component. Here $k=100$, $m=14$ and $\theta=0.421$ [rad].}
\label{fig:Qcoeff}
\end{figure}

We now take a closer look at the properties of the Fourier coefficients of $q_{k,m}(t;\theta)$, defined in \eqref{g j m q j m}. Numerical studies show that most of the energy of the set $\{Q_{k,m;\theta}[n]\}$ is concentrated around the direct current (DC) component. This behavior is typical to any choice of $k$, $m$ or $\theta$. An example for $k=100$, $m=14$ and $\theta=0.421$ [rad] is shown in Fig. \ref{fig:Qcoeff}.
This allows us to rewrite \eqref{c k m approx} as
\begin{equation}\label{c k m N1N2}
c_{k,m}^s\simeq\sum_{n=-N_1}^{N_2}\varphi_m^s[k-n]Q_{k,m;\theta}[n].
\end{equation}
The choice of $N_1$ and $N_2$ controls the approximation quality. Numerical studies show that $20$ most significant elements of $\{Q_{k,m;\theta}[n]\}$ contain, on average, more than $95\% $ of the entire energy irrespective of the choice of $k$, $m$ or $\theta$. Beamforming in frequency therefore is performed using $20$ most significant elements in $\{Q_{k,m;\theta}[n]\}$ throughout our work.

Denote by $\beta$, $|\beta | = B$, the set of Fourier coefficients of the detected signal that correspond to its bandwidth, namely, the values of $k$ for which $\varphi_m^s[k]$ is nonzero (or larger than a threshold).
Note that \eqref{c k m N1N2} implies, that the bandwidth of the beam, $\beta _{BF}$, will contain at most $(B+N_1+N_2)$ nonzero frequency components. To compute the elements in $\beta _{BF}$ all we need is the set $\beta$ for each one of the detected signals.  In a typical imaging setup $B$ is of order of hundreds of coefficients, while $N_1$ and $N_2$, defined by the decaying properties of $\{Q_{k,m;\theta}[n]\}$, are no larger than $10$. This implies that $B\gg N_i, i\in\{1,2\}$, so $B+N_1+N_2\approx B$. Hence, the bandwidth of the beam is approximately equal to the bandwidth of the detected signals.
In addition, it follows from \eqref{c k m N1N2} that in order to calculate an arbitrary subset $\mu\subset \beta _{BF}$ of size $M$ of Fourier coefficients of the beam, we need to know at most $(M+N_1+N_2)$ Fourier coefficients of each one of the detected signals $\varphi_m(t)$.
These properties of frequency domain beamforming will be used in order to reduce sampling rates.

Equations \eqref{fourier coeff of beamformed 2} and \eqref{c k m N1N2} provide a relationship between the Fourier series coefficients of the beam and the individual signals. We next derive a corresponding relationship between the DFT coefficients of the above signals, sampled at the beamforming rate $f_s$.
Denote by $N=\lfloor T\cdot f_s\rfloor$ the resulting number of samples.
Since $f_s$ is higher than the Nyquist rate of the detected signals, the relation between the DFT of length $N$ and the Fourier series coefficients of $\varphi_m(t)$ is given by \cite{oppenheim1999discrete}:
\begin{align}
\label{Fourier DFT detected signals}
\varphi_m^s [n] =\frac{1}{N} \left\{ \begin{array}{rl}
 & \hspace{-0.4cm} \varphi_m [n], \mbox{ ~~~~~~~~~$0\leq n \leq P$} \\
 & \hspace{-0.4cm} \varphi_m [N+n], \mbox{ ~~$-P\leq n < 0$} \\
  &\hspace{-0.4cm} 0, \mbox{~~~~~~~~~~~~~~ otherwise,}
       \end{array} \right.
\end{align}
where $\varphi_m [n]$ denote the DFT coefficients and $P$ denotes the index of the Fourier transform coefficient, corresponding to the highest frequency component.
We can use \eqref{Fourier DFT detected signals} to substitute Fourier series coefficients $\varphi_m^s[n]$ of $\varphi_m(t)$ in \eqref{c k m N1N2} by DFT coefficients $\varphi_m[n]$ of its sampled version. Plugging the result into \eqref{fourier coeff of beamformed 2}, we obtain a relationship between Fourier series coefficients of the beam and DFT coefficients of the sampled detected signals:
\begin{align}\label{fourierBF DFTdetected}
c_{k}^s \simeq & \frac{1}{MN} \sum_{m=1}^M \sum_{n=-N_1}^{k-\tilde{n}}\varphi_m[k-n]Q_{k,m;\theta}[n] \\ \nonumber
+ &\sum_{n=k-\tilde{n}+1}^{N_2}\varphi_m[k-n+N]Q_{k,m;\theta}[n]
\end{align}
for an appropriate choice of $\tilde{n}$.
Since $f_s$ is higher than the Nyquist rate of the beam as well, the DFT coefficients $c_k$ of its sampled version are given by an equation similar to \eqref{Fourier DFT detected signals}:
\begin{align}
\label{Fourier DFT BF signal}
c_k = N \left\{ \begin{array}{rl}
& \hspace{-0.4cm} c_k^s, \mbox{ ~~~~$0\leq k \leq P$} \\
& \hspace{-0.4cm} c_{k-N}^s, \mbox{ $N-P\leq k < N$} \\
& \hspace{-0.4cm} 0, \mbox{ ~~~~~~otherwise}.
       \end{array} \right.
\end{align}

Equations \eqref{fourierBF DFTdetected} and \eqref{Fourier DFT BF signal} provide the desired relationship between the DFT coefficients of the beam and the DFT coefficients of the detected signals.
Note that this relationship, obtained by a periodic shift and scaling of \eqref{c k m N1N2}, retains the important properties of the latter. 


Applying an IDFT on $\{c_k\}_{k=0}^{N-1}$ results in the beamformed signal in time. We can now proceed to standard image generation steps which include log-compression and interpolation.

\subsection{Simulations and Validation}
\label{ssec:simulations and validation}
\begin{figure}[htb]
\begin{minipage}[b]{0.48\linewidth}
  \centering
  \centerline{\includegraphics[width=5cm]{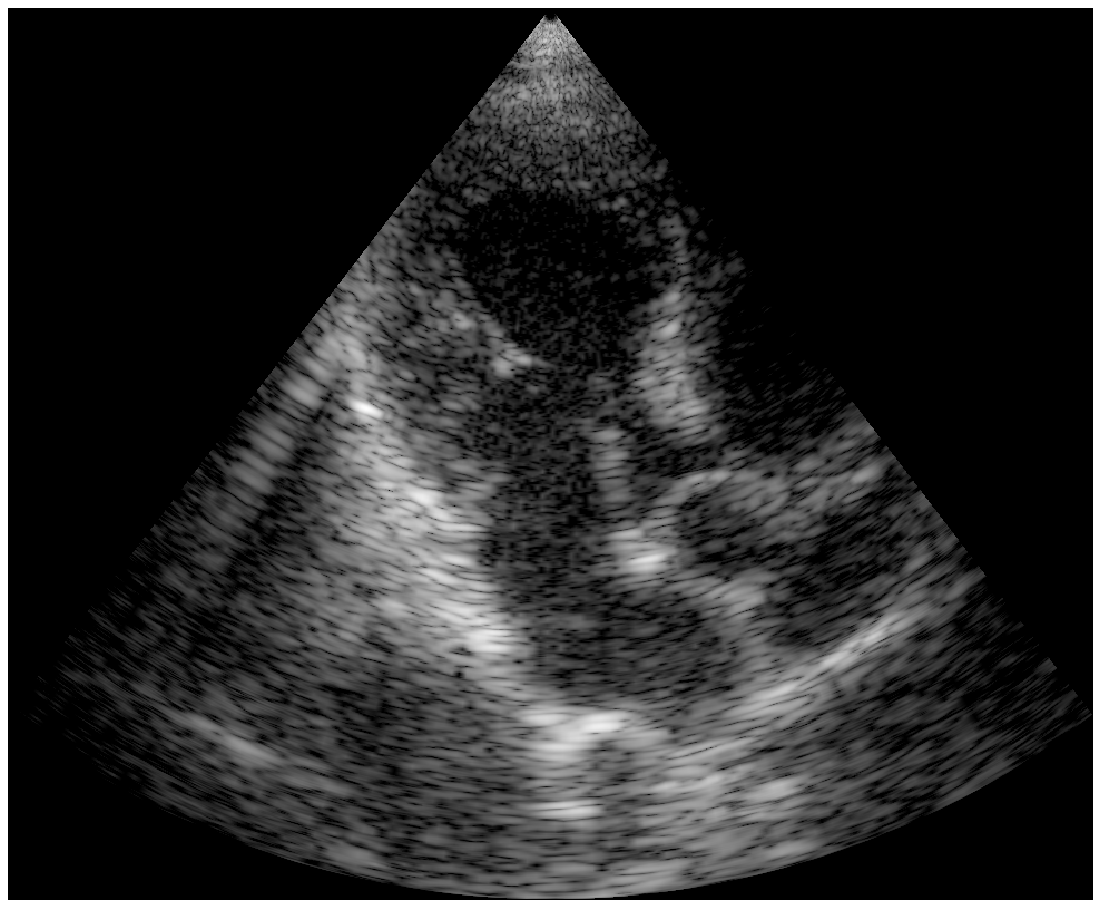}}
  \centerline{(a)}\medskip

\end{minipage}
\hfill
\begin{minipage}[b]{0.48\linewidth}
  \centering
  \centerline{\includegraphics[width=5cm]{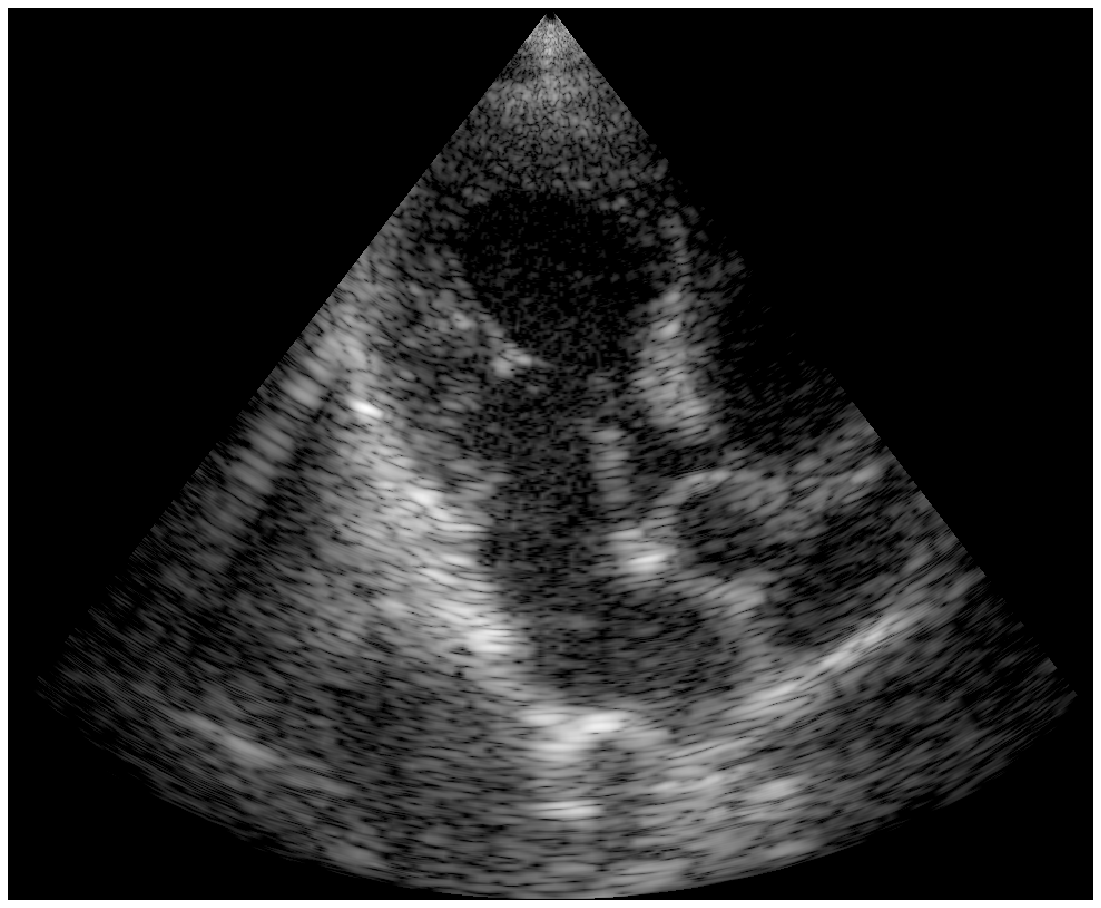}}
  \centerline{(b)}\medskip

\end{minipage}
\hfill

%
\caption{Cardiac images constructed with different beamforming techniques. (a) Time domain beamforming. (b) Frequency domain beamforming.  }
\label{fig:time vs freq beamformed}
\end{figure}

To demonstrate the equivalence of beamforming in time and frequency, we applied both methods on in vivo cardiac data yielding the images shown in Fig. \ref{fig:time vs freq beamformed}. The imaging setup is that described in Section \ref{ssec:rate requirements} with $f_s=16$ MHz. 
As can be readily seen, the images look identical.

Quantitative validation of the proposed method was performed  with respect to both one-dimensional beamformed signals and the resulting two-dimensional image. To compare the one-dimensional signals, we calculated the normalized root-mean-square error (NRMSE) between the signals obtained by beamforming in frequency and those obtained by standard beamforming in time. Both class of signals were compared after envelope detection, performed by a Hilbert transform in order to remove the carrier.
Denote by $ \Phi[n;\theta _j]$ the signal obtained by standard beamforming in direction $\theta _j, j=1,..., J$, and let $ \hat{\Phi}[n;\theta _j]$ denote the signal obtained by beamforming in frequency. The Hilbert transform is denoted by $H(\cdot)$. For the set of $J=120$ image lines, we define NRMSE as:
\begin{equation}\label{eq:NRMSE}
    NRMSE=\frac{1}{J} \frac{\sqrt{\frac{1}{N}\sum_{n=1}^N \left(H(\Phi[n;\theta _j])-H(\hat{\Phi}[n;\theta_j])\right)^2}}{H\left(\Phi[n;\theta _j]\right)_{max}-H\left(\Phi[n;\theta _j]\right)_{min}},
\end{equation}
where $H\left(\Phi[n;\theta _j]\right)_{max}$ and $H\left(\Phi[n;\theta _j]\right)_{min}$ denote the maximal and minimal values of the envelope of the beamformed signal in time.

Comparison of the resulting images was performed by calculating the structural similarity (SSIM) index \cite{wang2004image}, commonly used for measuring similarity between two images. The first line of Table \ref{table:timeVSfreq} summarizes the resulting values. These values verify that both 1D signals and the resulting image are extremely similar.

%

%
\section{Rate Reduction by Beamforming in Frequency}
\label{sec:rate reduction}
In the previous section we showed the equivalence of beamforming in time and frequency. We next demonstrate that beamforming in frequency allows to reduce the required number of samples of the individual signals. Reduction can be achieved in two different ways. First, we exploit the low effective bandwidth of ultrasound signals and bypass oversampling, dictated by digital implementation of beamforming in time. This allows to perform processing at the Nyquist rate, defined with respect to the effective bandwidth of the signal, which is impossible when beamforming is performed in time. As a second step, we show that further rate reduction is possible, when we take into account the FRI structure of the beamformed signal and use CS techniques for recovery.

In this section we address rate reduction, obtained by translation of the beamforming operator into the frequency domain. At this stage the structure of the beamformed signal is not taken into account.

\subsection{Exploiting Frequency Domain Relationship}
\label{ssec:red rate samp}
To reduce the rate, we exploit the relationship between the beam and the detected signals in the frequency domain given by \eqref{c k m N1N2}. In Section \ref{ssec:imlementation of BF in freq}, we showed that the bandwidth of the beam, $\beta_{BM}$, contains approximately $B$ nonzero frequency components, where $B$ is the effective bandwidth of the detected signals. In order to compute $\beta_{BM}$ we need a set $\beta$ of nonzero frequency components of each one of the detected signals.
This allows to exploit the low effective bandwidth of the detected signals and calculate only their nonzero DFT coefficients.
The ratio between the cardinality of the set $\beta$ and the overall number of samples $N$, required by standard beamforming rate $f_s$,  is dictated by the oversampling factor. As mentioned in Section \ref{ssec:rate requirements}, we define $f_s$ as $4$-$10$ times the bandpass bandwidth of the detected signal, leading to $B/N=1/4$-$1/10$. Assume that it is possible to obtain the required set $\beta$ using $B$ low-rate samples of the detected signal. In this case the ratio between $N$ and $B$ implies potential $4$-$10$ fold reduction in the required sampling rate.


\subsection{Reduced Rate Sampling}
\label{ssec:red rate samp}
We now address the following question: how do we obtain the required set $\beta$, corresponding to the effective bandpass bandwidth, using $B$ low-rate samples of each one of the detected signals?

\begin{figure}[htb]
\begin{minipage}[b]{1.0\linewidth}
  \centering
  \centerline{\includegraphics[width=4.5cm]{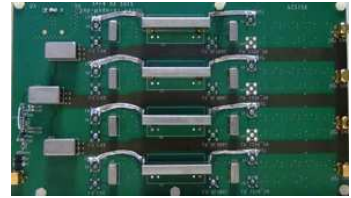}}
\end{minipage}
\caption{A  Xampling-based hardware prototype for sub-Nyquist sampling. The prototype computes low-rate samples of the input from which the set $\beta$ of DFT coefficients can be computed on the outputs. }
\label{fig:XamplingHardware}
\end{figure}

Note that sampling is performed in time, while our goal is to extract $B$ DFT coefficients. To this end, similarly to \cite{wagner2012compressed}, we can use the Xampling mechanism proposed in \cite{tur2011innovation}. A hardware Xampling prototype implemented by Baransky et al. in \cite{Baransky2012radar} is seen in Fig. \ref{fig:XamplingHardware}.

The Xampling scheme allows to obtain $B$ coefficients from $B$ point-wise samples of the detected signal filtered with an appropriate kernel $s^{\ast}(-t)$, designed according to the transmitted pulse-shape and the set $\beta$. The required DFT coefficients are equal to the DFT of the outputs, therefore, the number of samples taken at each individual element is equal to the number of DFT coefficients that we want to compute. Hence, when we compute all nonzero DFT coefficients of the detected signal, $4$-$10$ fold reduction in sampling rate is achieved without compromising image quality.

Having obtained the set $\beta$ of each one of the detected signals, we calculate the elements of $\beta_{BF}$ by low-rate frequency domain beamforming. Finally, we reconstruct the beamformed signal in time by performing an IDFT. Note that it is possible to pad the elements of $\beta_{BF}$ with an appropriate number of zeros to improve time resolution. In our experiments, in order to compare the proposed method with standard processing, we padded $\beta_{BF}$ with $N-B$ zeros, leading to the same sampling grid, used for high-rate beamforming in time.

\begin{figure}[htb]
\begin{minipage}[b]{0.48\linewidth}
  \centering
  \centerline{\includegraphics[width=5.0cm]{plots/images/timeBF_frame3}}
  \centerline{(a)}\medskip
\end{minipage}
\hfill
\begin{minipage}[b]{0.48\linewidth}
  \centering
  \centerline{\includegraphics[width=5.0cm]{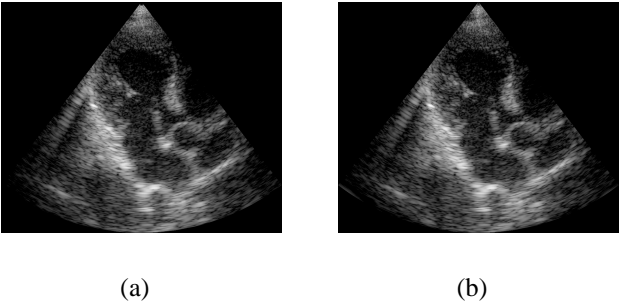}}
  \centerline{(b)}\medskip
\end{minipage}
\hfill
\caption{Cardiac images constructed with different beamforming techniques. (a) Time domain beamforming. (b) Frequency domain beamforming, obtained with 8 fold reduction in sampling rate.}
\label{fig:time vs freq beamformed reduced}
\end{figure}

Images obtained by the proposed method, using $416$ real-valued samples per image line to perform beamforming in frequency, and by standard beamforming, using $3360$ real-valued samples to perform beamforming in time, are shown in Fig. \ref{fig:time vs freq beamformed reduced}. Corresponding values of NRMSE and SSIM are reported in the second line of Table \ref{table:timeVSfreq}. These values validate close similarity between the two methods. However, in this case NRMSE is slightly higher, while SSIM is lower, compared to the values obtained in Section \ref{ssec:simulations and validation}. Note that these values depict similarity between the signals. Differences can therefore be explained by the following practical aspect.
When we obtain the set of all nonzero DFT coefficients of the beamformed signal, $\beta_{BF}$, all the signal energy is captured in the frequency domain. However, the signal obtained by beamforming in time, contains noise, which occupies the entire spectrum. When only the DFT coefficients within the bandwidth are computed in the frequency domain, the noise outside the bandwidth is effectively filtered out. In the signal obtained by standard beamforming, the noise is retained, reducing the similarity between the two signals.

\begin{figure}[htb]
\begin{minipage}[b]{1.0\linewidth}
  \centering
  \centerline{\includegraphics[width=8cm]{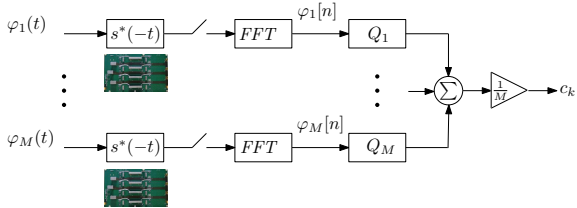}}
\end{minipage}
\caption{Fourier domain beamforming scheme. The block $Q_i$ represents averaging the DFT coefficients of the detected signals with weights $\{Q_{k,i;\theta}[n]\}$ according to \eqref{fourierBF DFTdetected} and \eqref{Fourier DFT BF signal}.}
\label{fig:BFsceme}
\end{figure}

The entire scheme, performing low-rate sampling and frequency domain beamforming, is depicted in Fig. \ref{fig:BFsceme}.
Signals $\left\{\varphi_m(t)\right\}_{m=1}^M$, detected at each transducer element, are filtered with an appropriate analog kernel $s^{\ast}(-t)$ and sampled at a low-rate, defined by the effective bandwidth of the transmitted pulse. Such a rate corresponds to the Nyquist rate of the baseband transmitted pulse. DFT coefficients of the detected signals are computed and beamforming is performed directly in frequency at a low-rate. This framework allows to bypass oversampling dictated by digital implementation of beamforming in time and to significantly reduce (up to $10$-fold) the resulting sampling rate.

\begin{table}[htdp] \caption{Quantitative validation of beamforming in frequency with respect to beamforming in time } \begin{center}
\begin{tabular}{|m{3.5cm}|c|c|c|c|}
  \hline
  \small{Method} & \small{NRMSE} &  \small{SSIM}\\ \hline \hline
  \small{Beamforming in frequency} & \small{0.0349} &\small{0.9684} \\ \hline
  \small{Beamforming in frequency, reduced rate sampling} & \small{ 0.0368} &\small{0.9603} \\ \hline
\end{tabular}
\end{center} \label{table:timeVSfreq}
\end{table}

\section{Further Reduction via Compressed Sensing}
\label{sec:CS}

We have shown that it is possible to reconstruct a beamformed signal perfectly from a set $\beta_{BF}$ of its nonzero DFT coefficients.
In this section we consider further reduction in sampling rate by taking only a subset $\mu\subset \beta_{BF}$, $|\mu|=M < B_{BF}=| \beta_{BF}|$, of nonzero DFT coefficients. In this case, as shown in Section \ref{ssec:imlementation of BF in freq}, $(M+N_1+N_2)$ frequency components of each one of the detected signals are required, leading to only $(M+N_1+N_2)$ samples in each channel.
The challenge now is to reconstruct a beamformed signal from such partial frequency data.

To this end we aim to use CS techniques, while exploiting the FRI structure of the beamformed signal. To formulate the recovery as a CS problem, we begin with a parametric representation of the beam.

\subsection{Parametric representation}
\label{ssec:param representation}
According to \cite{wagner2012compressed}, a beamformed signal obeys an FRI model, namely, it can be modeled as a sum of replicas of the known transmitted pulse, $h(t)$, with unknown amplitudes and delays:
\begin{equation}\label{beam FRI}
    \Phi(t;\theta)\simeq\sum_{l=1}^L\tilde{b}_l h(t-t_l),
\end{equation}
where $L$ is the number of scattering elements in direction $\theta$, $\{\tilde{b}_l\}_{l=1}^L$ are the unknown amplitudes of the reflections and $\{t_l\}_{l=1}^L$ denote the times at which the reflection from the $l$th element arrived at the reference receiver $m_0$. Since the transmitted pulse is known, such a signal is completely defined by $2L$ unknown parameters, the amplitudes and the delays.

We can rewrite this model accordingly by
sampling both sides of \eqref{beam FRI} at the beamforming rate $f_s$ and quantizing the unknown delays $\{t_l\}_{l=1}^L$ with quantization step $1/f_s$, such that $t_l=q_l/f_s, q_l\in\mathbb{Z}$ and $N=\lfloor T\cdot f_s\rfloor$:
\begin{equation}\label{beam FRI n}
    \Phi[n;\theta]\simeq\sum_{l=1}^L\tilde{b}_l h[n-q_l]=\sum_{l=0}^{N-1} b_l h[n-l],
\end{equation}
where
%
\begin{align}
\label{tilde b - b}
b_l = \left\{ \begin{array}{rl}
 &\hspace{-0.5cm}\tilde{b}_l, ~\mbox{ if $l=q_l$} \\
 &\hspace{-0.5cm} 0, ~~\mbox{ otherwise}.
       \end{array} \right.
\end{align}
Calculating the DFT of both sides of \eqref{beam FRI n} leads to the following expression for the DFT coefficients $c_k$:
\begin{align}
\label{DFT BF}
c_k=\sum_{n=0}^{N-1}\Phi[n;\theta]e^{-i \frac{2\pi}{N}k n}=h_k\sum_{l=0}^{N-1} b_l e^{-i \frac{2\pi}{N}k l},
\end{align}
where $h_k$ is the DFT coefficient of $h[n]$, the transmitted pulse sampled at rate $f_s$. We conclude that recovering $\Phi[n;\theta]$ is equivalent to determining $b_l$, $0\leq l\leq N-1$ in \eqref{DFT BF}.

We now recast the problem in vector-matrix notation.
Defining an $M$-length measurement vector $\mathbf{c}$ with $k$th entry $c_k, k\in\mu$, we can rewrite \eqref{DFT BF} as follows:
\begin{equation}\label{eq:c=HDb}
\mathbf{c}=\mathbf{H} \mathbf{D} \mathbf{b} = \mathbf{A} \mathbf{b} ,
\end{equation}
where $\mathbf{H}$ is an $M\times M$ diagonal matrix with $h_k$ as its $k$th entry, $\mathbf{D}$ is an $M\times N$ matrix formed by taking the set $\mu$ of rows from an $N\times N$ DFT matrix, and $\mathbf{b}$ is a length-$N$ vector with $l$th entry $b_l$.

Our goal is to determine $\mathbf{b}$ from $\mathbf{c}$.  We next discuss and compare possible recovery approaches.

\subsection{Prior Work}
\label{ssec:prior work}

As mentioned in Section \ref{ssec:param representation}, the signal of interest is completely defined by $L$ the unknown delays and amplitudes. Hence, a possible approach is to extract those values from the available set $\mu$ of DFT coefficients. To this end, we can view \eqref{eq:c=HDb} as a complex sinusoid problem. For $M\geq2L$ it can be solved using standard spectral analysis methods such as matrix pencil \cite{sarkar1995using} or annihilating filter \cite{stoica1997introduction}. Rate reduction is achieved when $2L<<N$, where $N$ is the number of samples dictated by the standard beamforming rate. 

In the presence of moderate to high noise levels, the unknown parameters can be extracted more efficiently using a CS approach, as was shown in \cite{wagner2012compressed}.
Note that \eqref{eq:c=HDb} is an underdetermined system of linear equations which has infinitely many solutions, since $\mathbf{A}$ is an $M\times N$ matrix with $M\ll N$. The solution set can be narrowed down to a single value by exploiting the structure of the unknown vector $\mathbf{b}$. In the CS framework it is assumed that the vector of interest is reasonably sparse, whether in the standard coordinate basis or with respect to some other basis.

The regularization introduced in \cite{wagner2012compressed}, relies on the assumption that the coefficient vector $\mathbf{b}$ is $L$-sparse. The formulation in \eqref{eq:c=HDb} then has a form of a classic CS problem, where the goal is to reconstruct an $N$-dimensional $L$-sparse vector $\mathbf{b}$ from its projection onto $K$ orthogonal rows captured by the measurement matrix $\mathbf{A}$. This problem can be solved using numerous CS techniques, when $\mathbf{A}$ satisfies well-known properties such as restricted isometry (RIP) or coherence \cite{eldar2012compressed}.

In our case, $\mathbf{A}$, defined in \eqref{eq:c=HDb}, is formed by taking $K$ scaled rows from an $N\times N$ DFT matrix. It can be shown that by choosing $K\geq C L(\log N)^4$ rows uniformly at random for some positive constant $C$, the measurement matrix $\mathbf{A}$ obeys the RIP with high probability \cite{rudelson2008sparse}. In order for this approach to be beneficial it is important to assume that $L<<N$.
Since random frequency sampling is not practical from a hardware prospective, it is possible instead to sample a number of frequency bands, distributed randomly throughout the spectrum \cite{Baransky2012radar}. This approach is implemented in the board of Fig. \ref{fig:XamplingHardware}.

A typical beamformed ultrasound signal is comprised of a relatively small number of strong reflections, corresponding to strong perturbations in the tissue, and many weaker scattered echoes, originated from microscopic changes in acoustic impedance of the tissue.
The framework proposed in \cite{wagner2012compressed} aims to recover only strong reflectors in the tissue and treat weak echoes as noise. Hence, the vector of interest $\mathbf{b}$ is indeed $L$-sparse with $L<<N$.
To recover $\mathbf{b}$, Wagner et al. consider the following optimization problem:
\begin{equation}\label{eq:l0 minimization}
    \min_{\mathbf{b}}\|\mathbf{b}\|_0  \textrm{~~~subject to~~~} \|\mathbf{A}\mathbf{b}-\mathbf{c}\|_2\leq\varepsilon,
\end{equation}
where $\varepsilon$ is an appropriate noise level, and approximate its solution using orthogonal matching pursuit (OMP) \cite{tropp2007signal}.

A significant drawback of this method is its inability to restore weak reflectors. In the context of this approach they are treated as noise and are disregarded by the signal model. As a result, the speckle - granular pattern that can be seen in Fig. \ref{fig:time vs freq beamformed} - is lost. This severely degrades the value of the resulting images since information carried by speckle is of major importance in many medical imaging modalities. For example, in cardiac imaging, speckle tracking tools allow to analyze the motion of heart tissues and to track effectively myocardial deformations \cite{notomi2005measurement}, \cite {suffoletto2006novel}.

\subsection{Alternative Approach}
\label{ssec:alternative approach}

Fortunately, with a small conceptual change of model, we can restore the entire signal, namely, recover both strong reflectors and weak scattered echoes.

As mentioned above, a beamformed ultrasound signal is comprised of a relatively small number of strong reflections and many scattered echoes, that are on average two orders of magnitude weaker. It is, therefore, natural to assume that the coefficient vector $\mathbf{b}$, defined in \eqref{eq:c=HDb}, is compressible or approximately sparse, but not exactly sparse. This property of $\mathbf{b}$ can be captured by using the $\textit{l}_1$ norm,
leading to the optimization problem:
\begin{equation}\label{eq:l1 minimization}
    \min_{\mathbf{b}}\|\mathbf{b}\|_1  \textrm{~~~subject to~~~} \|\mathbf{A}\mathbf{b}-\mathbf{c}\|_2\leq\varepsilon.
\end{equation}
Problem \eqref{eq:l1 minimization} can be solved using second-order methods such as interior point methods \cite{candes2007l1}, \cite{grant2008cvx} or first-order methods, based on iterative shrinkage ideas \cite{beck2009fast}, \cite{hale2007fixed}.

We emphasize that although it is common to view \eqref{eq:l1 minimization} as a convex relaxation of \eqref{eq:l0 minimization}, in our case such a substitution is crucial. It allows to capture the structure of the signal and to boost the performance of sub-Nyquist processing, as will be shown next, through several examples.

\section {Simulations and Results}
\label{sec:simulations and results}
In this section we examine the performance of low-rate frequency-domain beamforming using $\textit{l}_1$ optimization and compare it to the previously proposed $\textit{l}_0$ optimization based method. This is done by applying both methods to stored RF data, acquired from a healthy volunteer. We then integrate our method into a stand alone ultrasound machine and show that such processing is feasible in practice using real hardware.
\subsection{Simulations on In Vivo Cardiac Data}
\label{cardiac simulations}

\begin{figure*}[!t]
\centering
\subfloat[]{\includegraphics[width=6 cm]{plots/images/timeBF_frame3}%
\label{fig_first_case}}
\hfil
\subfloat[]{\includegraphics[width=6 cm]{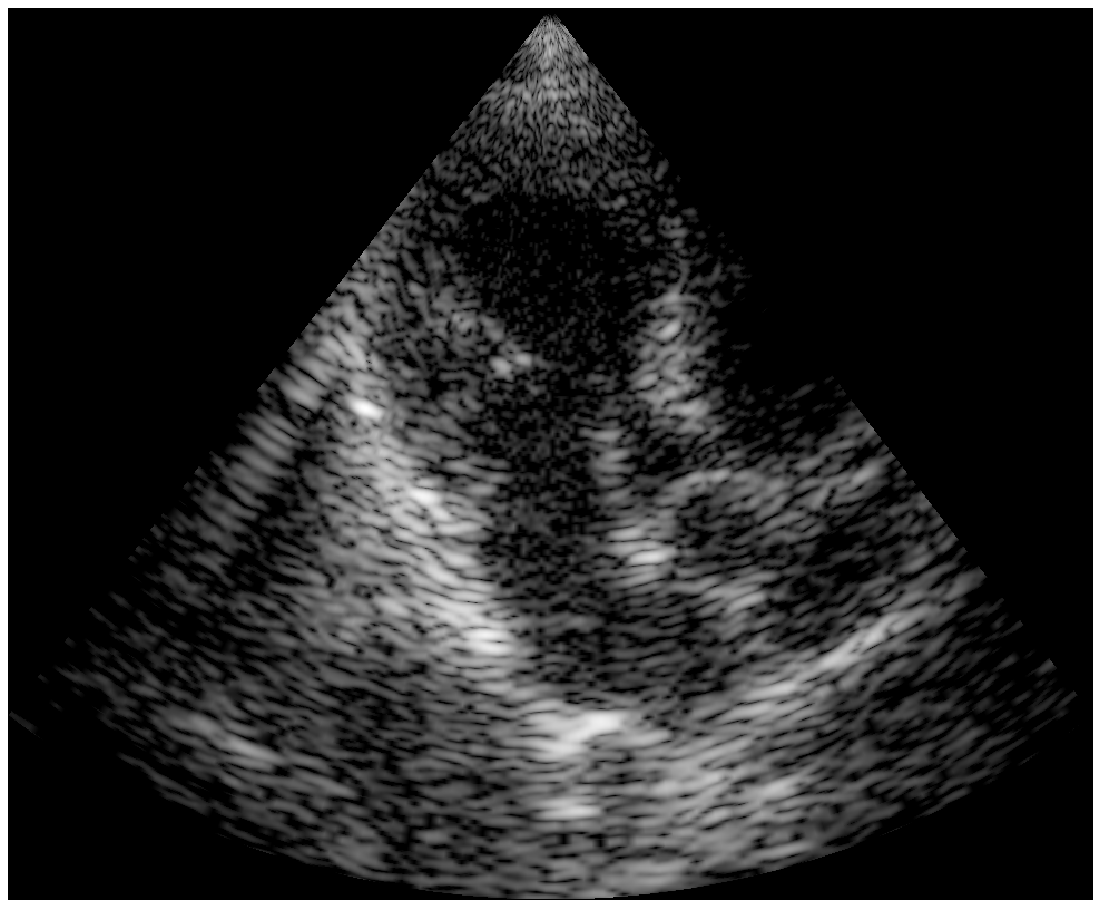}%
\label{fig_second_case}}
\hfil
\subfloat[]{\includegraphics[width=6 cm]{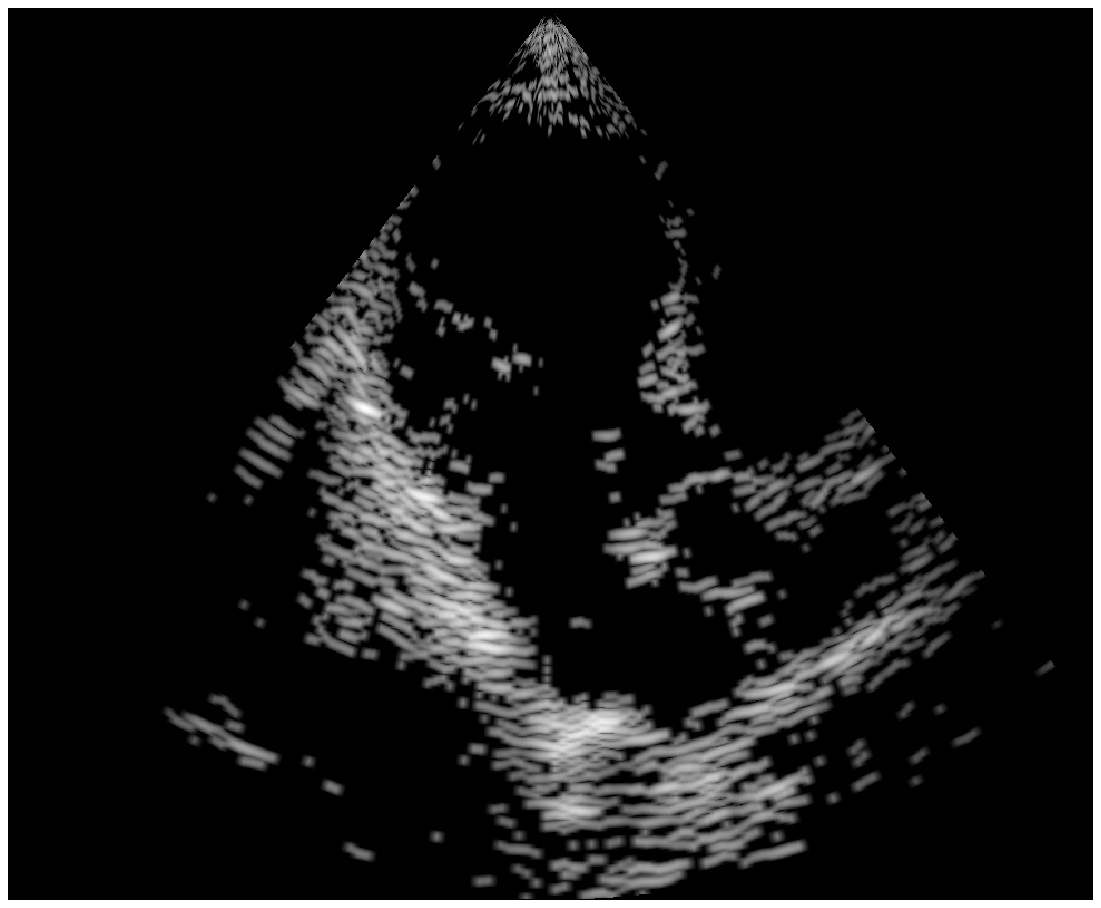}%
\label{fig_third_case}} \\
\centering
\subfloat[]{\includegraphics[width=6 cm]{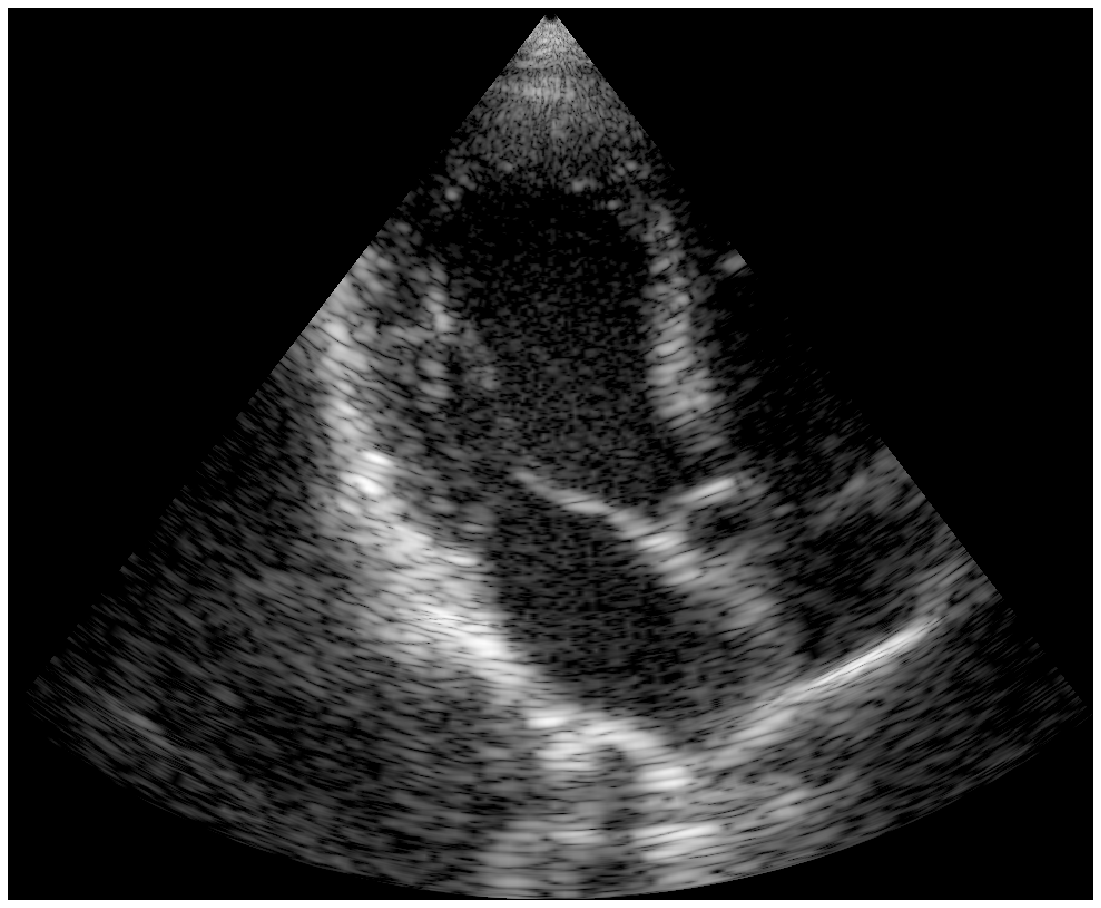}%
\label{fig_first_case2}}
\hfil
\subfloat[]{\includegraphics[width=6 cm]{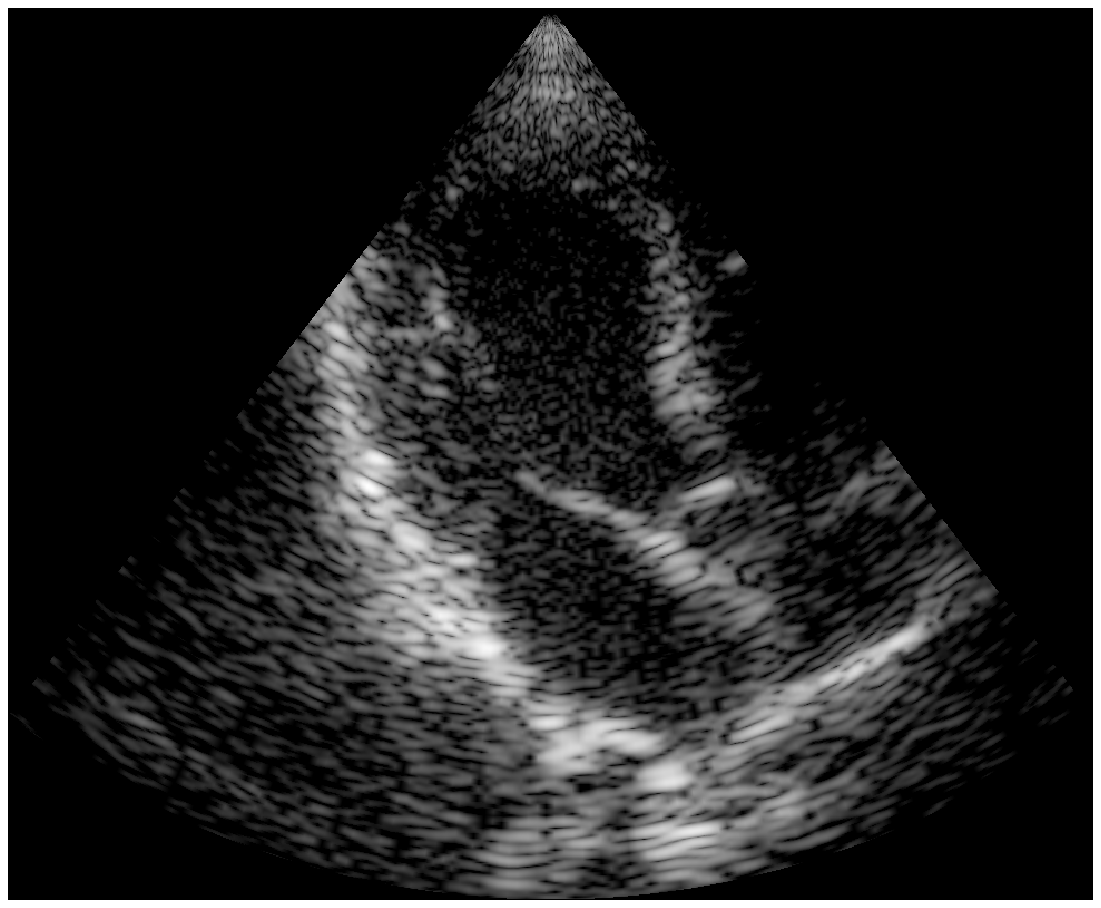}%
\label{fig_second_case2}}
\hfil
\subfloat[]{\includegraphics[width=6 cm]{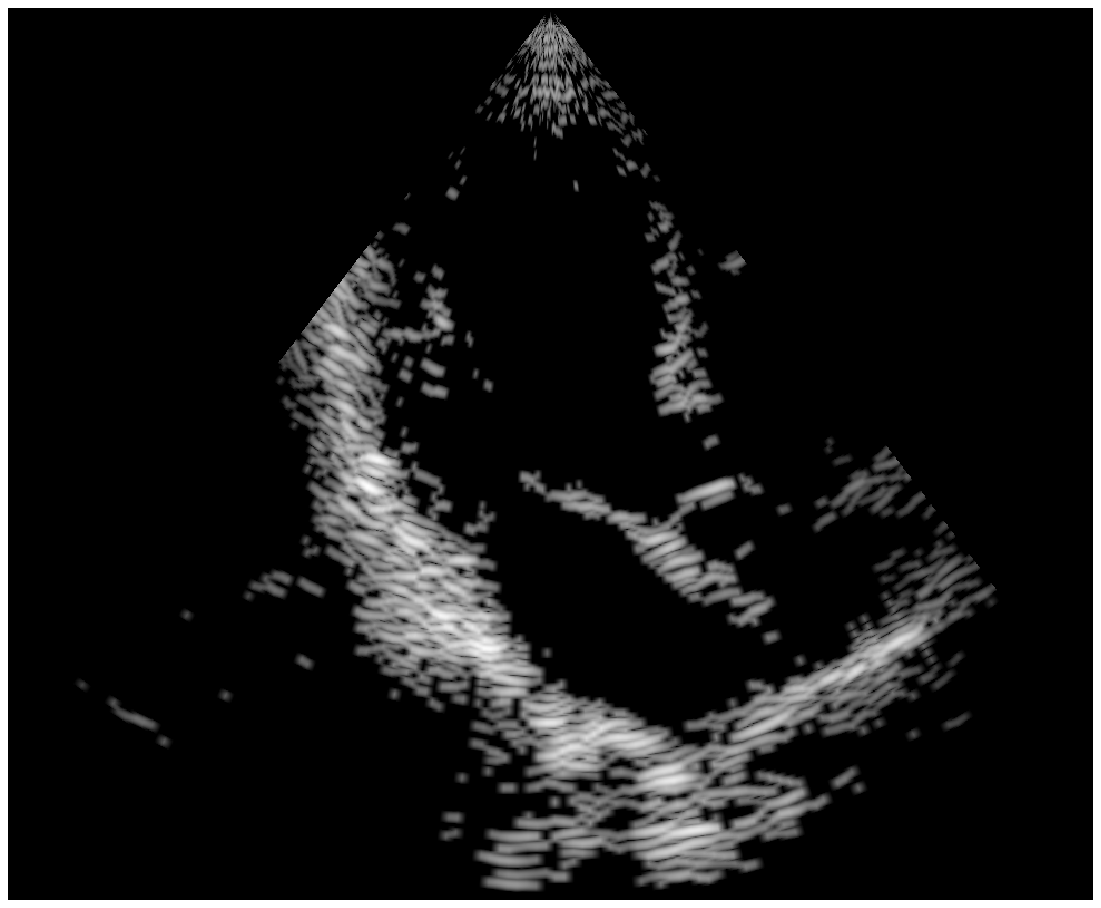}%
\label{fig_third_case2}}
\caption{Simulation results. The first row, (a)-(c), corresponds to frame 1, the second row, (d)-(f), corresponds to frame 2. (a), (d) Time domain beamforming. (b), (e) Frequency domain beamforming,  $\textit{l}_1$  optimization solution. (c),(f) Frequency domain beamforming,  $\textit{l}_0$  optimization solution.}.
\label{fig:frame3}
\end{figure*}


To demonstrate low-rate beamforming in frequency and evaluate the impact of rate reduction on image quality, we applied our method on in vivo cardiac data. The data acquisition setup is described in Section \ref{ssec:rate requirements} with $f_s=16$ MHz, leading to $3360$ real valued samples. To perform beamforming in frequency we used a subset $\mu$ of $100$ DFT coefficients, which can be obtained from $120$ real-valued samples by the proposed Xampling scheme. This implies $28$ fold reduction in sampling
and $14$ fold reduction in processing
rate compared to standard beamforming, which requires $3360$ real-valued samples for this particular imaging setup. The difference between the sampling and processing rates stems from the complex nature of DFT coefficients. Having computed the DFT coefficients of the beamformed signal, we obtain its parametric representation by solving \eqref{eq:l1 minimization}. To this end we used the NESTA algorithm \cite{becker2011nesta}. This fast and accurate first-order method, based on the work of Nesterov \cite{nesterov2005smooth}, is shown to be highly suitable for solving \eqref{eq:l1 minimization}, when the signal of interest is compressible with high dynamic range, which is particulary true for ultrasound imaging. An additional advantage of NESTA is that it does not depend on fine tuning of numerous controlling parameters. A single smoothing parameter, $\mu$, needs to be selected based on a trade-off between the accuracy of the algorithm and its speed of convergence. This parameter was chosen empirically to yield optimal performance with respect to image quality.

The resulting images, corresponding to two different frames, are shown in Figs. \ref{fig:frame3} (b) and (e). Although the images are not identical to those obtained by standard beamforming (Figs. \ref{fig:frame3} (a) and (d)), it can be easily seen that $\textit{l}_1$ optimization, based on the assumption that the signal of interest is compressible, allows to reconstruct both strong reflectors and speckle.
Table \ref{table:timeVSfreqSubNyquist} reports corresponding values of NRMSE and SSIM. Although the quantitative values are reduced compared to those obtained in Sec. \ref{ssec:red rate samp}, important information, e.g. the thickness of the heart wall and the valves, as well as the speckle pattern, essential for tracking tools, are preserved.

We would like to emphasize, that the values of NRMSE and SSIM are provided in order to give a sense of performance of the proposed method. In practice, unfortunately, there are no established quantitative measures for the quality of ultrasound images. Validation is typically performed visually by sonographers, radiologists and physicians. Furthermore, our approach inherently reduces noise so that high similarity with beamforming in time may not necessarily be advantageous.

\begin{table}[htdp] \caption{Quantitative validation of beamforming in frequency at sub-Nyquist rate} \begin{center}
\begin{tabular}{|m{1.5cm}|c|c|c|c|}
  \hline
  \small{Method} & \small{NRMSE} &  \small{SSIM}\\ \hline \hline
  \small{Frame 1} & \small{ 0.0682} &\small{ 0.7017} \\ \hline
  \small{Frame 2} & \small{ 0.0587} &\small{ 0.6812} \\ \hline
 \end{tabular}
\end{center} \label{table:timeVSfreqSubNyquist}
\end{table}

To compare the proposed method with the previously developed  $\textit{l}_0$ optimization based approach, we solved \eqref{eq:l0 minimization} with OMP, while assuming $L=25$ strong reflectors in each direction $\theta$. Resulting images, shown in Figs. \ref{fig:frame3}(c) and (f), depict the strong reflectors, observed in Fig. \ref{fig:frame3}(a) and (b), while the speckle is completely lost, degrading the overall image.

%

\subsection{Implementation on Stand Alone Imaging System}
\label{implementaiton on system}

\begin{figure}[htb]
\begin{minipage}[b]{1.0\linewidth}
  \centering
  \centerline{\includegraphics[width=9.5cm]{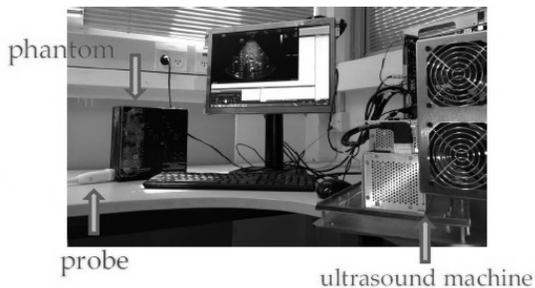}}
\end{minipage}
\vspace{-1.4cm}
\caption{Lab setup: Ultrasound system, probe and cardiac phantom.}
\label{fig:labsetup}
\end{figure}

\begin{figure}[htb]
\begin{minipage}[b]{1.0\linewidth}
  \centering
 \vspace{.1cm}
  \centerline{\includegraphics[width=8.0cm]{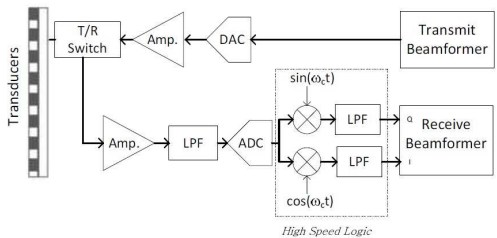}}
  \vspace{.1cm}
\end{minipage}
\caption{Transmit and receive front-end of a medical ultrasound system.}
\label{fig:SystemInTime}
\end{figure}

As a next step we implemented low-rate frequency domain beamforming on an ultrasound imaging system \cite{eilam2013cloud}. The lab setup used for implementation and testing is shown in Fig. \ref{fig:labsetup} and includes a state of the art GE ultrasound machine, a phantom and an ultrasound scanning probe.
In our study we used a breadboard ultrasonic scanner with 64 acquisition channels. The radiated depth $r=15.7$ cm and speed of sound $c=1540$ m/sec yield a signal of duration $T=2r/c\simeq204$ $\mu$sec. The acquired signal is characterized by a narrow bandpass bandwidth of $1.77$ MHz, centered at a carrier frequency $f_0\approx3.4$ MHz. The signals are sampled at the rate of $50$ MHz and then are digitally demodulated and down-sampled to the demodulated processing rate of $f_p\approx2.94$ MHz, resulting in $1224$ real-valued samples per transducer element. Linear interpolation is then applied in order to improve beamforming resolution, leading to $2448$ real valued samples.
Fig. \ref{fig:SystemInTime} presents a schematic block diagram of the transmit and receive front-end of the medical ultrasound system being used.

\begin{figure}[htb]
\begin{minipage}[b]{1.0\linewidth}
  \centering
  \centerline{\includegraphics[width=9.5cm]{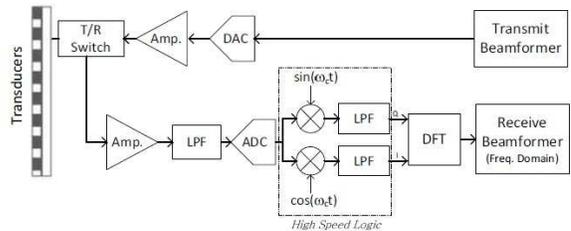}}
\end{minipage}
\caption{Transmit and receive paths of a medical ultrasound system with beamforming in the frequency domain.}
\label{fig:SystemInFreq1}
\end{figure}

\begin{figure}[htb]
\begin{minipage}[b]{0.48\linewidth}
  \centering
  \centerline{\includegraphics[width=5.0cm]{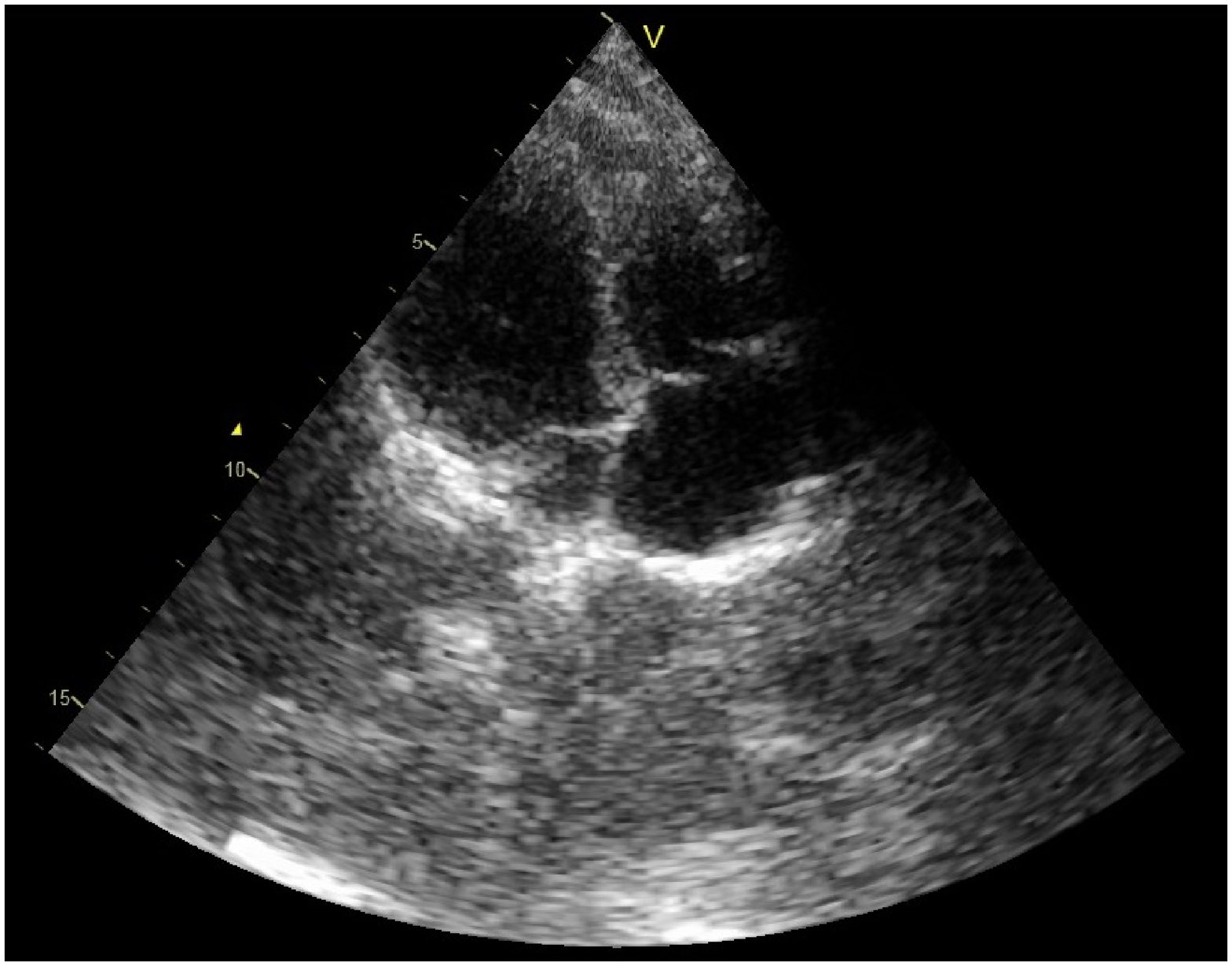}}
  \centerline{(a)}\medskip
\end{minipage}
\hfill
\begin{minipage}[b]{0.48\linewidth}
  \centering
  \centerline{\includegraphics[width=5.1cm]{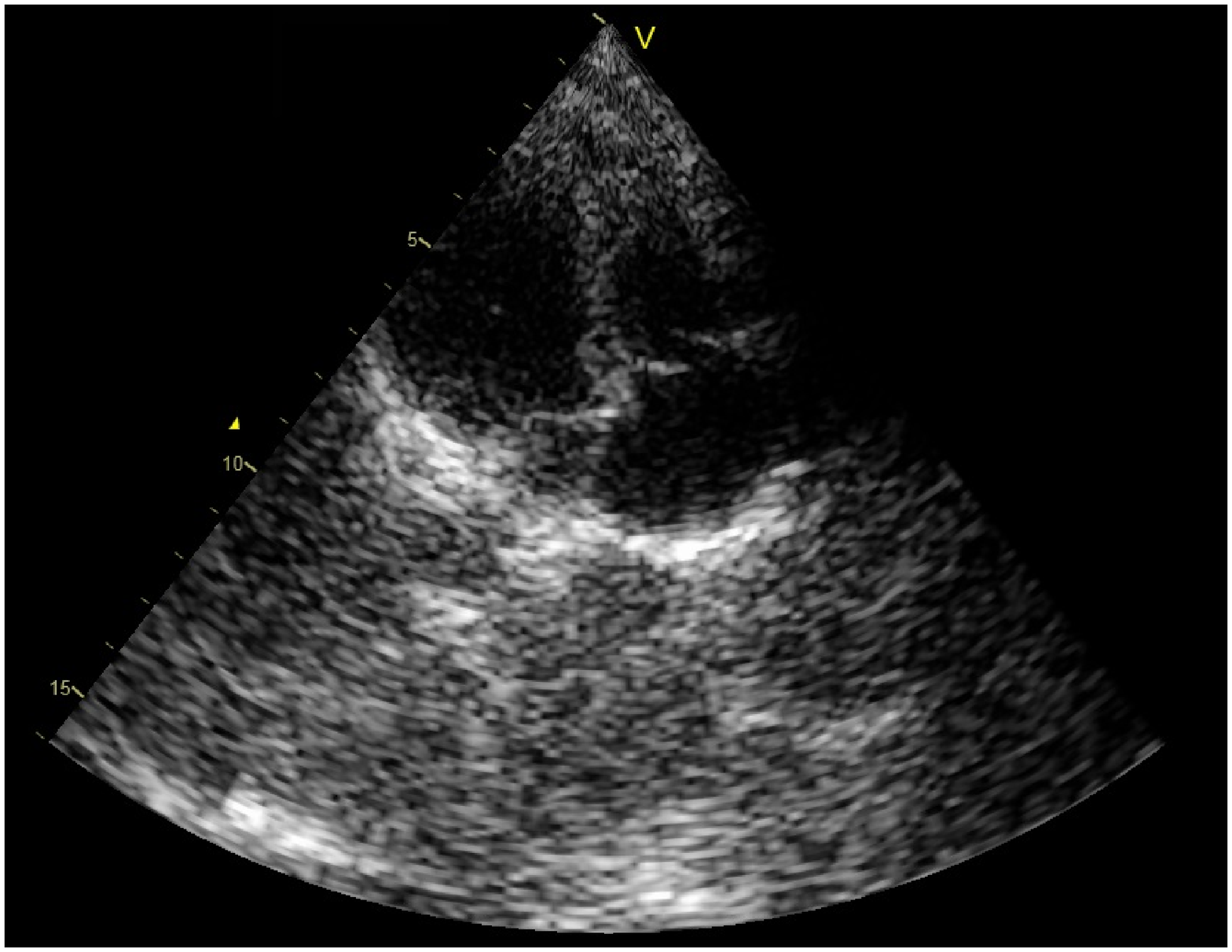}}
  \centerline{(b)}\medskip
\end{minipage}
\hfill
\caption{Cardiac images obtained by demo system. (a) Time domain beamforming. (b) Frequency domain beamforming, obtained with $10$ fold reduction in processing rate.}
\label{fig:demoResults}
\end{figure}

At this point of our work, as illustrated in Fig. \ref{fig:SystemInFreq1}, in-phase and quadrature components of the detected signals were used to obtain the desired set of their DFT coefficients.
Using this set, beamforming in frequency was performed according to \eqref{fourierBF DFTdetected} and \eqref{Fourier DFT BF signal}, yielding the DFT coefficients of the beamformed signal. In this setup the sampling rate remained unchanged, but  frequency domain beamforming was performed at a low rate.
In our experiments we computed $100$ DFT coefficients of the beamformed signal, using $120$ DFT coefficients of each one of the detected signals. This corresponds to $240$ real-valued samples used for beamforming in frequency. The number of samples required by demodulated processing rate is $2448$. Hence, beamforming in frequency is performed at a rate corresponding to $240/2448\approx1/10$ of the demodulated processing rate.
Images obtained by low-rate beamforming in frequency and standard time-domain beamforming are presented in Fig. \ref{fig:demoResults}. As can be readily seen, we are able to retain sufficient image quality despite the significant reduction in processing rate.
%


Our implementation was done on a state-of-the-art system, sampling each channel at a high rate. Data and processing rate reduction took place following DFT, in the frequency domain. However, by implementing the Xampling scheme described in Section \ref{ssec:red rate samp}, the set of $120$ DFT coefficients of the detected signals, required for frequency domain beamforming, can be obtained directly from only $120$ real-valued low rate samples.

\section{Conclusion}
\label{sec:conclusions}

In this work we extended the compressed beamforming framework, proposed in \cite{wagner2012compressed}, to a general concept of beamforming in frequency, dual to standard time domain beamforming.
We have shown that when performed directly in frequency, beamforming does not require oversampling, essential for its digital implementation in time. Hence, $4$-$10$ fold reduction in sampling rate is achieved by the translation of beamforming into the frequency domain, without compromising image quality and without involving any additional assumptions on the signal.

Further reduction in sampling rate is obtained, when only a portion of the beam's bandwidth is used. In this case the detected signals are sampled at a sub-Nyquist rate, leading to up to $28$ fold reduction in sampling rate. In order to reconstruct the beamformed signal from such partial frequency data, we rely on the fact that the beamformed signal obeys an FRI model and  use CS techniques. To improve the performance of sub-Nyquist processing and avoid the loss of speckle information, we assumed that the coefficient vector is compressible. This assumption allows to capture both strong reflections, corresponding to large perturbations in the tissue, and much weaker scattered echoes, originating from microscopic changes in acoustic impedance of the tissue.


Finally, we implemented our frequency domain beamforming on a stand alone ultrasound machine. Low-rate processing is performed on the data obtained in real-time by scanning a heart with a $64$ element probe. The proposed approach allows for $10$ fold rate reduction with respect to the lowest processing rates that are achievable today.

Our results prove that the concept of sub-Nyquist processing is feasible for medical ultrasound, leading to the potential of considerable reduction in future ultrasound machines size, power consumption and cost.

\section*{Acknowledgment}
The authors would like to thank GE Healthcare Haifa and in particular Dr. Arcady Kempinski for providing the imaging system and for many helpful discussions. They are grateful to Alon Eilam for his assistance with the implementation of the proposed method on an ultrasound machine.

\bibliographystyle{IEEEtran}
\bibliography{IEEEfull,general}

\end{document}